\documentclass[a4paper,aps,preprint,nofootinbib,floatfix,superscriptaddress,prd]{revtex4-2}
\usepackage{graphicx,color}
\usepackage{booktabs}
\usepackage{amsmath,amssymb,mathtools}
\usepackage{dsfont}
\usepackage[pdfusetitle,pdfencoding=auto]{hyperref}

\begin{document}

\title{Confronting the prediction of leptonic Dirac CP-violating phase with experiments}

\author{Yang Hwan Ahn}
\email{axionahn@naver.com}
\author{Sin Kyu Kang}
\email{skkang@seoultech.ac.kr}
\author{Raymundo Ramos}
\email{rayramosang@gmail.com}
\affiliation{%
School of Liberal Arts, Seoul National University of Science and Technology,
232 Gongneung-ro, Nowon-gu, Seoul, 01811, Korea
}
\author{Morimitsu Tanimoto}
\email{tanimoto@muse.sc.niigata-u.ac.jp}
\affiliation{%
Department of Physics, Niigata University, Ikarashi 2-8050, Niigata 950-2181, Japan
}

\date{\today}

\begin{abstract}
    We update and improve past efforts to predict the leptonic Dirac CP-violating phase
    with models that predict perturbatively modified tribimaximal or bimaximal mixing.
    Simple perturbations are applied to both mixing patterns
    in the form of rotations between two sectors.
    By translating these perturbed mixing matrices to the standard parameterization
    for the neutrino mixing matrix we derive relations between the Dirac CP-phase
    and the oscillation angles.
    We use these relations together with current experimental results
    to constrain the allowed range for the CP-phase
    and determine its probability density.
    Furthermore, we elaborate on the prospects for future experiments
    probing on the perturbations considered in this work.
    We present a model with $A_4$ modular symmetry
    that is consistent with one of the described perturbed scenarios
    and successfully predicts current oscillation parameter data.
\end{abstract}


\maketitle
\newpage

\section{Introduction}

With the discovery of the Higgs scalar by the LHC in 2012
the standard model (SM) of particle physics took the seat
as the most predictive high energy theory so far.
While further experimental efforts keep giving results that are mostly consistent with the SM,
one of its sectors has, since long ago, given the best motivation
for physics beyond the SM\@: Neutrinos.
First proposed as a way to fix conservation laws in beta decays,
they have had an eventful history,
while they went from massless
to having tiny masses
and changing flavor---oscillate---while travelling
due to mixing between flavor states.
The first evidence of neutrino oscillations was reported in 1998 \cite{Super-Kamiokande:1998kpq}.
Neutrino oscillations were firmly established in 2001
using solar neutrinos~\cite{SNO:2001kpb},
and, since then, experiments regularly close in on their oscillation pattern
and the mass differences responsible of these oscillations.
Fast forwarding to 2011 and 2012,
the Double Chooz~\cite{DoubleChooz:2011ymz} and Daya Bay~\cite{DayaBay:2012fng} experiments
measured $\theta_{13} \neq 0$ with enough precision
to open the possibility of a Dirac type CP-violating phase
in the mixing of the leptonic sector of the SM
described through the Pontecorvo-Maki-Nakagawa-Sakata (PMNS) matrix~\cite{Pontecorvo:1957cp,Maki:1962mu}.

The usual approach to extend the SM to include neutrino masses and mixing
employs a discrete flavor symmetry at a very high energy.
After the spontaneous breaking of this symmetry at lower energies,
residual symmetries remain in the charged and neutral leptons mass masses,
thus, resulting in particular mixing patterns in the PMNS matrix, $U^\text{PMNS}$.
Before the measurement of the reactor angle, $\theta_{13} \neq 0$,
models that predicted no mixing between first and third family were popular,
in particular models that predicted two maximal oscillations
popularly known as bimaximal (BM) mixing~\cite{Vissani:1997pa,Barger:1998ta,Baltz:1998ey,Fritzsch:1998xs,Mohapatra:1998bp,Kang:1998gs,Altarelli:2009gn}
and, as more data accumulated,
other works appeared suggesting maximal mixing of two and three families,
known as tribimaximal (TBM) mixing~\cite{Harrison:2002er,Harrison:2002kp,Xing:2006xa,He:2003rm}.
Naturally, after the measurement of a non-zero reactor angle,
the exact BM and TBM mixing patterns were ruled out.
In more complicated formulations,
these patterns can be considered the result of residual symmetries
that need to be broken by perturbations that permit the appearance of a non-zero reactor angle.
Interestingly, this type of formulations often result in relationships between oscillation parameters
that allow an estimation of the size of the Dirac type leptonic CP-violating phase.
This is the idea that was developed in Ref.~\cite{Kang:2014mka}
as well as in other several works~\cite{Kang:2000sb,Fukugita:2001rk,Giunti:2002pp,Xing:2002sw,Guo:2003cc,Petcov:2004rk,Ge:2011ih,Marzocca:2011dh,He:2011kn,Shimizu:2014ria,Kang:2015xfa,Kang:2017uqi,Delgadillo:2018tza,Kang:2018txu}.
In the present work we attempt to follow up on the scenarios explored in Ref.~\cite{Kang:2014mka}
and extend the analysis to probability densities for the CP-violating phase
based on currently allowed ranges for oscillation parameters from experiments.
Moreover, we simulate the effects of the constraints from these scenarios
to estimate their chances of survival
in three long-baseline experiments that may be operative in the near future.
We complete this work by showing how one of these scenarios
can be realized in a flavor model of neutrino masses and mixing.

The rest of the paper is laid out as follows:
In Sec.~\ref{sec:pertTBM} we introduce the perturbations to TBM mixing
that will be used along the rest of this work
and their constraints on oscillation parameters.
In Sec.~\ref{sec:pertScenarios} we present probability densities
related to the CP-violating phase considering constraints
from the cases of Sec.~\ref{sec:pertTBM}\@.
In Sec.~\ref{sec:pertBM} we describe the perturbed scenarios
applied to BM mixing and comment on the effects of current experimental
constraints.
In Sec.~\ref{sec:prospects} we present the prospects of future experiments
expected to constrain the scenarios considered here.
In Sec.~\ref{sec:modularA4} we construct a model using $A_4$ modular symmetry
that is consistent with one of the perturbed scenarios
and expand on its properties.
Finally, in Sec.~\ref{sec:conclusion} we discuss the most relevant details
of this work and conclude.

\section{Perturbative modifications to tribimaximal mixing}
\label{sec:pertTBM}

Let us begin by recalling the form of the exact TBM mixing matrix~\cite{Harrison:2002er}
\begin{equation}
\label{eq:exactTBM}
    U^{\rm TBM}_0 = \begin{pmatrix}
         \sqrt{\frac{2}{3}} &  \frac{1}{\sqrt{3}} & 0 \\
        -\sqrt{\frac{1}{6}} &  \frac{1}{\sqrt{3}} & \frac{1}{\sqrt{2}} \\
         \sqrt{\frac{1}{6}} & -\frac{1}{\sqrt{3}} & \frac{1}{\sqrt{2}}
    \end{pmatrix}.
\end{equation}
As mentioned before, this mixing matrix form
has been the motivation for a great number of models
that attempt to predict the neutrino oscillation parameters
employing discrete symmetries.
It is this sort of pattern with a vanishing 1-3 matrix element
that were ruled out by the measurement of the non-zero reactor angle $\theta_{13}$.
In this paper we will consider the following
minimal perturbations to the TBM mixing matrix
\begin{equation}
\label{eq:cases}
    V=
    \begin{cases}
            U_{0}^{\rm TBM} U_{23}(\theta, \phi) \quad \mbox{(Case A)},\\
            U_{0}^{\rm TBM} U_{13}(\theta, \phi) \quad \mbox{(Case B)},\\
            U^{\dagger}_{12}(\theta, \phi) U_{0}^{\rm TBM} \quad \mbox{(Case C)},\\
            U^{\dagger}_{13}(\theta, \phi) U_{0}^{\rm TBM} \quad \mbox{(Case D)}.
    \end{cases}
\end{equation}
where the $U_{ij}(\theta,\phi)$ matrices are given by
\begin{align}
    \label{eq:UPerturb}
U_{12}(\theta,\phi) & = \begin{pmatrix}
    \cos\theta           & -\sin\theta e^{i\phi} & 0 \\
    \sin\theta e^{-i\phi} & \cos\theta             & 0 \\
    0                    & 0                      & 1
    \end{pmatrix},\\
U_{13}(\theta,\phi) & = \begin{pmatrix}
    \cos\theta           & 0 & -\sin\theta e^{i\phi}  \\
    0                    & 1 & 0 \\
    \sin\theta e^{-i\phi} & 0 & \cos\theta
    \end{pmatrix},\\
U_{23}(\theta,\phi) & = \begin{pmatrix}
    1 & 0                    & 0 \\
    0 & \cos\theta           & -\sin\theta e^{i\phi} \\
    0 & \sin\theta e^{-i\phi} & \cos\theta 
    \end{pmatrix}.
\end{align}
Finding the equivalence between the mixing matrix $V$ of each case and the $U^\text{PMNS}$
can be done elementwise with $V_{ij}\exp(\alpha_i + \beta_j) = U^\text{PMNS}_{ij}\exp(\varphi_j)$.

The exact TBM pattern of Eq.~\eqref{eq:exactTBM}
can be regarded as result of residual symmetries
in the charged lepton and neutrino sectors
from a flavor model defined at a higher energy.
In this case, the mixing matrices of cases A and B in Eq.~\eqref{eq:cases}
can be considered the consequence of additional effects
that break these residual symmetries
on the planes (2,3) and (1,3)
in the side of the neutrino sector, respectively.
Similarly, cases C and D would break the residual symmetries
on the side of the charged leptons
on the planes (1,2) and (1,3), respectively.
Note that cases A and C were also studied in
Refs.~\cite{Rodejohann:2012cf,Marzocca:2013cr,Xing:2014zka}.
To simplify the notation,
we will be using the shorthand $s_{ij} = \sin\theta_{ij}$ and $c_{ij} = \cos\theta_{ij}$ in the rest of the paper.

\section{CP-violating phase from perturbative scenarios}
\label{sec:pertScenarios}

\begin{table}[tb]
    \setlength\tabcolsep{0.2cm}
    \begin{tabular}{lcccc}
        \toprule
        Parameter                            & Best fit$\pm 1\sigma$ (NO) & $3\sigma$ range (NO) & Best fit$\pm 1\sigma$ (IO) & $3\sigma$ range (IO) \\
        \cmidrule(r){1-1} \cmidrule(lr){2-3} \cmidrule(lr){4-5}
        $\sin^2\theta_\text{12}$             & $0.304\pm 0.012$           & [0.269, 0.343]       & $0.304^{+0.013}_{-0.012}$  & [0.269, 0.343]       \\
        $\sin^2\theta_\text{13}$ [$10^{-2}$] & $2.246\pm 0.062$           & [2.060, 2.435]       & $2.241^{+0.074}_{-0.062}$  & [2.055, 2.457]       \\
        $\sin^2\theta_\text{23}$             & $0.450^{+0.019}_{-0.016}$  & [0.408, 0.603]       & $0.570^{+0.016}_{-0.022}$  & [0.410, 0.613]       \\
        $\delta_\text{CP}$ [deg]             & $230^{+36}_{-25}$          & [144, 350]           & $278^{+22}_{-30}$          & [194, 345]           \\
        $\Delta m_{21}^2$ [$10^{-5}$ eV$^2$] & $7.42^{+0.21}_{-0.20}$     & [6.82, 8.04]         & $7.42^{+0.21}_{-0.20}$     & [6.82, 8.04]         \\
        $\Delta m_{3k}^2$ [$10^{-3}$ eV$^2$] & $2.510\pm 0.027$           & [2.430, 2.593]       & $-2.490^{+0.26}_{-0.28}$   & $[-2.574, -2.410]$   \\
        \bottomrule
    \end{tabular}
    \caption{\label{tab:nufit5_1_SK}%
        Oscillation parameters for three neutrino flavors
        as reported in NuFIT 5.1~\cite{Gonzalez-Garcia:2021dve}
        for normal ordering (NO) ($\Delta m_{3k}^2 = \Delta m_{31}^2$)
        and inverted ordering (IO) ($\Delta m_{3k}^2 = \Delta m_{32}^2$),
        including the tabulated $\chi^2$ data from Super-Kamiokande.}
\end{table}

One of the most relevant points of enabling a non-zero $\theta_{13}$
is opening up the possibility of having a Dirac-type CP-violating phase
in the PMNS mixing matrix.
Due to the features of the cases mentioned in Eq.~\eqref{eq:cases}
it is possible to relate either $\theta_{12}$ (cases A and B) or $\theta_{23}$ (cases C and D)
with the reactor angle $\theta_{13}$
and, lastly, to relate the $\delta_{CP}$ phase to the pair of free mixing angles.
This is achieved by identifying the parameterizations that result from Eq.~\eqref{eq:cases}
with the standard PDG parameterization of the PMNS matrix.
In this way, in Ref.~\cite{Kang:2014mka} the following relations
between oscillation parameters were worked out:
\begin{align}
    \label{eq:A12delCP}
    \text{A:}&\qquad s_{12}^2 = 1 - \frac{2}{3(1 - s_{13}^2)}, &&
        \cos\delta_\text{CP} = \frac{5 s_{13}^2 - 1}{\eta_{23}s_{13}\sqrt{2 - 6 s_{13}^2}}, \\
    \label{eq:B12delCP}
    \text{B:}&\qquad s_{12}^2 = \frac{1}{3(1 - s_{13}^2)}, &&
        \cos\delta_\text{CP} = \frac{2 - 4 s_{13}^2}{\eta_{23}s_{13}\sqrt{2 - 3 s_{13}^2}}, \\
    \label{eq:C12delCP}
    \text{C:}&\qquad s_{23}^2 = 1 - \frac{1}{2(1 - s_{13}^2)}, &&
        \cos\delta_\text{CP} = \frac{s_{13}^2 - (1 - 3 s_{12}^2)(1 - 3 s_{13}^2)}{3 s_{13}\xi\sqrt{1 - 2 s_{13}^2}}, \\
    \label{eq:D12delCP}
    \text{D:}&\qquad s_{23}^2 = \frac{1}{2(1 - s_{13}^2)}, &&
        \cos\delta_\text{CP} = \frac{(1 - 3 s_{12}^2)(1 - 3 s_{13}^2) - s_{13}^2}{3 s_{13}\xi\sqrt{1 - 2 s_{13}^2}},
\end{align}
where $\eta_{23} = 2\tan 2\theta_{23}$ and $\xi = \sin 2 \theta_{12}$.
Considering the form of the matrices of Eq.~\eqref{eq:UPerturb}
we can write the following expressions for the other oscillation parameters
in terms of the angle $\theta$ and the phase $\phi$:
\begin{align}
    \label{eq:A1323}
    \text{A:}&\quad s_{13}^2 = \frac{\sin^2\theta}{3}, &&
        s^2_{23} = \frac{3 - \sin^2\theta + \sqrt{6} \sin 2\theta\cos\phi}{6 - 2 \sin^2\theta}, \\
    \label{eq:B1323}
    \text{B:}&\quad s_{13}^2 = \frac{2\sin^2\theta}{3}, &&
        s^2_{23} = \frac{3 - 2\sin^2\theta + \sqrt{3} \sin 2\theta\cos\phi}{6 - 4 \sin^2\theta}, \\
    \label{eq:CD1312}
    \text{C, D:}&\quad s^2_{13} = \frac{\sin^2\theta}{2}, &&
        s_{12}^2 = \frac{2}{3}\left(\frac{1 - \sin 2\theta \cos\phi}{2 - \sin^2\theta}\right),
\end{align}
Note that, for every case,
there is a relationship between $\theta_{13}$ and $\theta$,
consistent with the idea that the matrices in Eq.~\eqref{eq:UPerturb}
are perturbations that deviate $\theta_{13}$ from zero.
Using Eqs.~\eqref{eq:A12delCP} to~\eqref{eq:CD1312},
other noteworthy consequences of these perturbations include
that for case A $s_{12}^2<1/3$,
while for case B $s_{12}^2>1/3$,
resulting in case B not being able to reproduce the current best fit value
for this oscillation parameter.
For cases C and D we obtain $s_{23}^2 < 1/2$ and $s_{23}^2 > 1/2$, respectively,
meaning that whenever the octant of $\theta_{23}$ is resolved
at least one of these two cases will be ruled out.

\subsection{Probability densities of \texorpdfstring{$\cos\delta_\text{CP}$}{cos delta CP}}
\label{sec:probDensCosDelta}

Using the expressions in Eqs.~\eqref{eq:A12delCP} to~\eqref{eq:D12delCP}
and the measured oscillation parameters
from NuFIT 5.1 global fit~\cite{Gonzalez-Garcia:2021dve},
we can calculate probability densities
for the predictions of the $\delta_\text{CP}$ phase
in every scenario.
The process for calculating these densities follows Ref.~\cite{Everett:2019idp}.
There are three facts that simplify the process in the present case:
\begin{enumerate}
    \item
        Eqs.~\eqref{eq:A1323} to~\eqref{eq:CD1312} imply an upper bound on $s_{13}^2$
        that is well above the acceptable experimental range
        and, thus, has no relevant effect in this analysis.
    \item
        With the same equations, the values we can get for $s_{23}^2$ in cases A and B
        are not particularly limited by specific values of $s_{13}^2$
        in the range of interest from the global fit,
        therefore, we can consider $s_{13}^2$ independent of $s_{23}^2$.
    \item
        Using input values around 3$\sigma$ range
        for $s_{13}^2$ and $s_{23}^2$ in cases A and B
        predicts only physical values for $\cos\delta_\text{CP}$.
\end{enumerate}
Points 2 and 3 above are also true for cases C and D replacing $s_{23}^2$ by $s_{12}^2$.
Considering these details, we can calculate the probability density for $\cos\delta_\text{CP}$
directly using the probability densities of $s_{13}^2$, $s_{23}^2$ and $s_{12}^2$.
The integral that we have to perform
to calculate the probability density
at some particular value $z$ of $\cos\delta_\text{CP}$
is given by
\begin{align}
    \label{eq:PcosdAB}
    P^\text{(A,B)}_{\cos\delta_\text{CP}} (z) & = \int dx\, dy\,\delta(f_\text{A,B}(x, y) - z) P_{s_{13}^2}(x)P_{s_{23}^2}(y),\\
    \label{eq:PcosdCD}
    P^\text{(C,D)}_{\cos\delta_\text{CP}} (z) & = \int dx\, dw\,\delta(f_\text{C,D}(x, w) - z) P_{s_{13}^2}(x)P_{s_{12}^2}(w),
\end{align}
where $w$, $x$, $y$ represent values of $s_{12}^2$, $s_{13}^2$, $s_{23}^2$, respectively,
that we have to integrate over.
The functions $f_j$, with $j\in\{\text{A,B,C,D}\}$,
represent $\cos\delta_\text{CP}$ for each case
and the delta function ensures that the integration is performed
over a line where $\cos\delta_\text{CP} = z$.
Independently of the three points enumerated before,
the probability densities $P_{s_{ij}^2}$
can be any normalized function where the values of $z$
are well defined in the integration intervals.
Note that, in general, one of the two probability densities in each integral
should be a conditional probability distribution
dependent on the input of the other,
however, given point 2 above,
we are considering both distributions in each integral as independent.

\begin{figure}[tb]
    \includegraphics{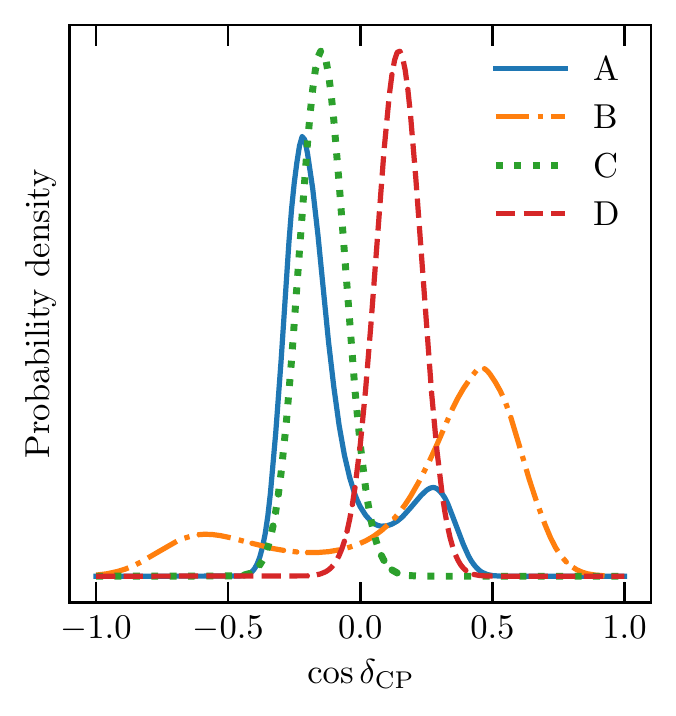}%
    \includegraphics{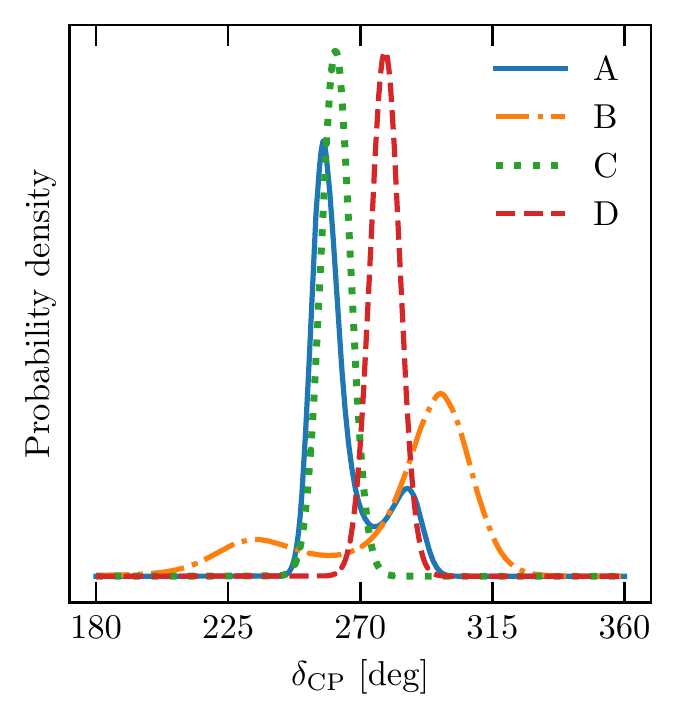}
    \caption{\label{fig:densCosDeltaCP}%
        Probability densities for $\cos\delta_\text{CP}$ (left)
        and $\delta_\text{CP}$ (right)
        using experimental results for normal ordering.
        The probability densities were obtained
        using the method detailed in Sec.~\ref{sec:probDensCosDelta}.}
\end{figure}

For this work, we are interested in using Eqs.~\eqref{eq:PcosdAB} and~\eqref{eq:PcosdCD}
to calculate $\cos\delta_\text{CP}$ probability densities
from currently observed oscillation parameters.
For this purpose we use the $\chi^2$ tables
provided by the NuFIT collaboration available on their website~\cite{nufitwebsite}.
The data used corresponds to the normal (NO) and inverted (IO) ordering results that include Super-Kamiokande's
tabulated $\chi^2$ data (lower part of Table 3 in Ref.~\cite{Gonzalez-Garcia:2021dve}),
these have been collected in Table~\ref{tab:nufit5_1_SK} for convenience.
The $\chi^2$ values are used to construct
probability densities of the form $P(\alpha) = N\exp(-\chi^2(\alpha)/2)$,
where $N=(\int d\alpha \exp(-\chi^2(\alpha)/2))^{-1}$ ensures that the probability density
integrates to one.
The probability densities obtained with the method described above
are shown in Fig.~\ref{fig:densCosDeltaCP}
for $\cos\delta_\text{CP}$ and $\delta_\text{CP}$.
For cases A (blue line), C (green dotted line) and D (red dashed line),
the prediction for $\cos\delta_\text{CP}$ lies inside the $[-0.5, 0.5]$ range,
with case D mostly positive
while case C is mostly negative.
Case A has a more spread distribution
but the most probable range for $\cos\delta_\text{CP}$
is predicted close to $-0.25$.
The probability that corresponds to case B (orange dash-dotted line)
is distributed along nearly all the $[-1, 1]$ range
with its highest peak around 0.5.
For the CP-violating phase $\delta_\text{CP}$
this means that A, C and D are close to $90^\circ$ or $270^\circ$.
Note that the right side of Fig.~\ref{fig:densCosDeltaCP}
only shows the range $[180^\circ, 360^\circ]$,
which is currently favoured by observations.
The range $(0,180^\circ)$ is just a mirror image of said figure.

\begin{figure}[tb]
    \includegraphics{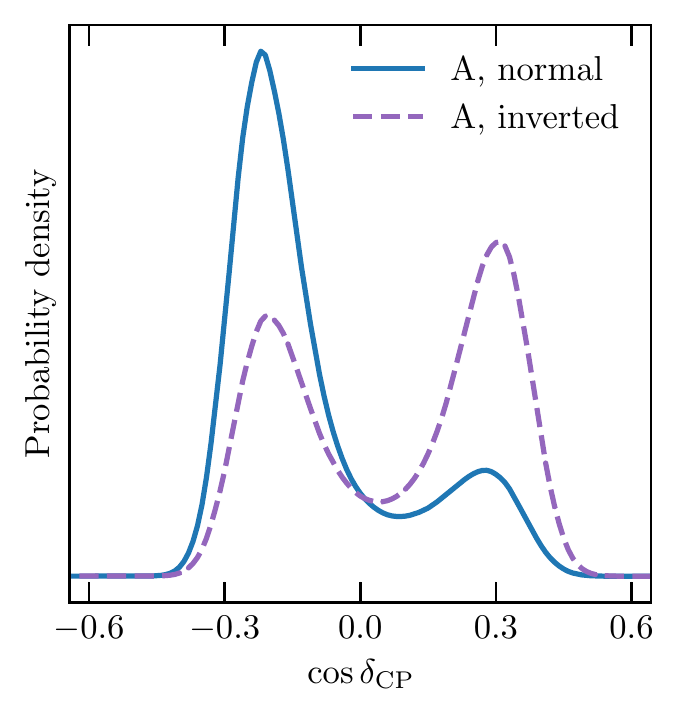}%
    \includegraphics{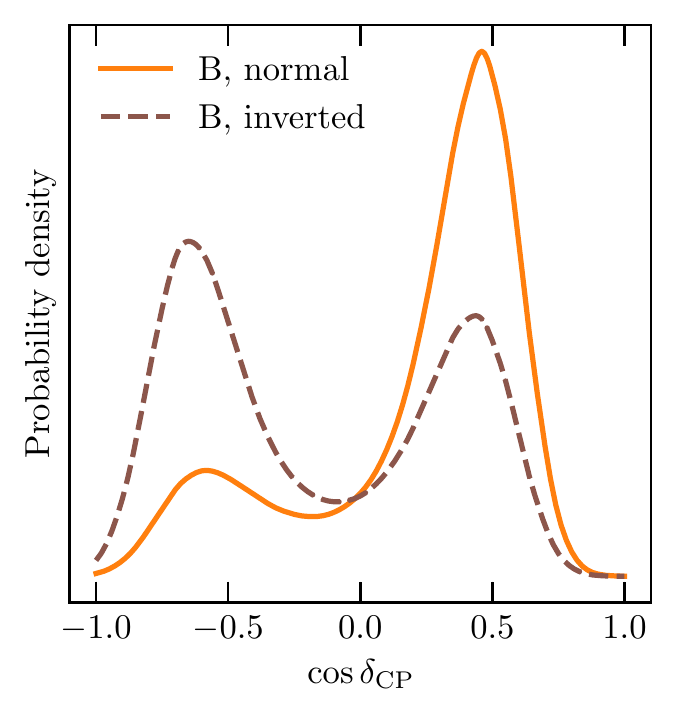}
    \caption{\label{fig:densCosDeltaNvsI}%
        Differences in the distribution of $\cos\delta_\text{CP}$
        for cases A (left) and B (right)
        when considering data for NO (solid) and IO (dashed).
        The probability densities were obtained
        using the method detailed in Sec.~\ref{sec:probDensCosDelta}.}
\end{figure}

The changes in the distributions of $\cos\delta_\text{CP}$
from considering NO or IO data
from Table~\ref{tab:nufit5_1_SK} is mostly related
to changes in central values and $\chi^2$ projections.
Cases C and D, which depend on $s_{13}^2$ and $s_{12}^2$,
do not change notably between using NO or IO data.
However, in the case of A and B,
due to the significant change in the $\chi^2$ projection of $s_{23}^2$,
the distribution of $\cos\delta_\text{CP}$ changes
to display two more leveled peaks with the higher peak changing side in both cases.
This can be see in detail in Fig.~\ref{fig:densCosDeltaNvsI},
where we can see that for case A in IO (left pane, dashed line) the highest peak changes to $\sim 0.3$
while for case B using IO data (right pane, dashed line) the highest peak moves to $\sim -0.65$.
These changes can be interpreted
as the delta CP phase $\delta_\text{CP}$ in case A
changing from $256^\circ$ in NO to $288^\circ$ in IO,
while for case B it changes from $297^\circ$ in NO to $230^\circ$ in IO.

\section{Perturbative modifications to bimaximal mixing}
\label{sec:pertBM}

In the same way we modified the TBM mixing case in Sec.~\ref{sec:pertTBM},
we can apply perturbations to the very well known
BM mixing~\cite{Vissani:1997pa,Barger:1998ta,Baltz:1998ey,Fritzsch:1998xs,Mohapatra:1998bp,Kang:1998gs,Altarelli:2009gn}.
The exact form of the BM mixing matrix is given by
\begin{equation}
\label{eq:exactBM}
    U^{\rm BM}_0 = \begin{pmatrix}
         \sqrt{\frac{1}{2}} &  \frac{1}{\sqrt{2}} & 0 \\
        -\frac{1}{2} &  \frac{1}{2} & \frac{1}{\sqrt{2}} \\
         \frac{1}{2} & -\frac{1}{2} & \frac{1}{\sqrt{2}}
    \end{pmatrix}.
\end{equation}
From here, perturbations proceed identically as for the TBM case.
We can define the following scenarios
\begin{equation}
\label{eq:BMcases}
    V=
    \begin{cases}
            U^{\dagger}_{12}(\theta, \phi) U_{0}^{\rm BM} \quad \mbox{(Case E)},\\
            U^{\dagger}_{13}(\theta, \phi) U_{0}^{\rm BM} \quad \mbox{(Case F)},\\
            U_{0}^{\rm BM} U_{23}(\theta, \phi) \quad \mbox{(Case G)},\\
            U_{0}^{\rm BM} U_{13}(\theta, \phi) \quad \mbox{(Case H)}.\\
    \end{cases}
\end{equation}
with the $U_{ij}(\theta,\phi)$ matrices given in Eq.~\eqref{eq:UPerturb}.
Cases G and H were considered ruled out by experimental data above 3$\sigma$
when they were studied on Ref.~\cite{Kang:2014mka}.
For cases E and F the expressions for $s_{23}^2$ are identical to those
of cases C and D, respectively.
The relationships between mixing angles and CP-violating phase
are given by
\begin{align}
    \label{eq:E12delCP}
    \text{E:}&\qquad \cos\delta_\text{CP} = \frac{3 s_{13}^2 - 1}{\eta_{12}s_{13}\sqrt{1 - 2 s_{13}^2}}, \\
    \label{eq:F12delCP}
    \text{F:}&\qquad \cos\delta_\text{CP} =  \frac{1 - 3 s_{13}^2}{\eta_{12}s_{13}\sqrt{1 - 2 s_{13}^2}},
\end{align}
where $\eta_{12} = 2\tan 2\theta_{12}$.

\begin{figure}[tb]
    \includegraphics{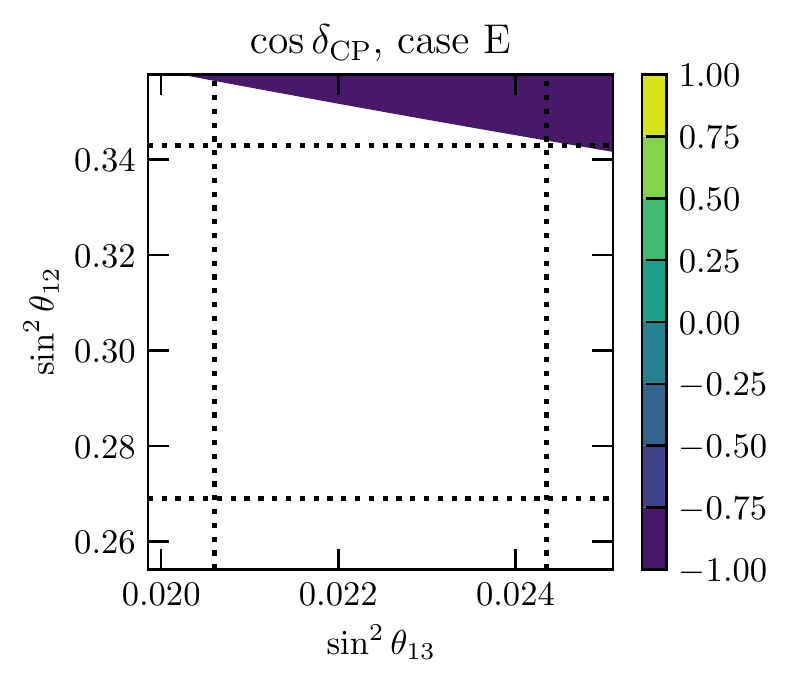}%
    \includegraphics{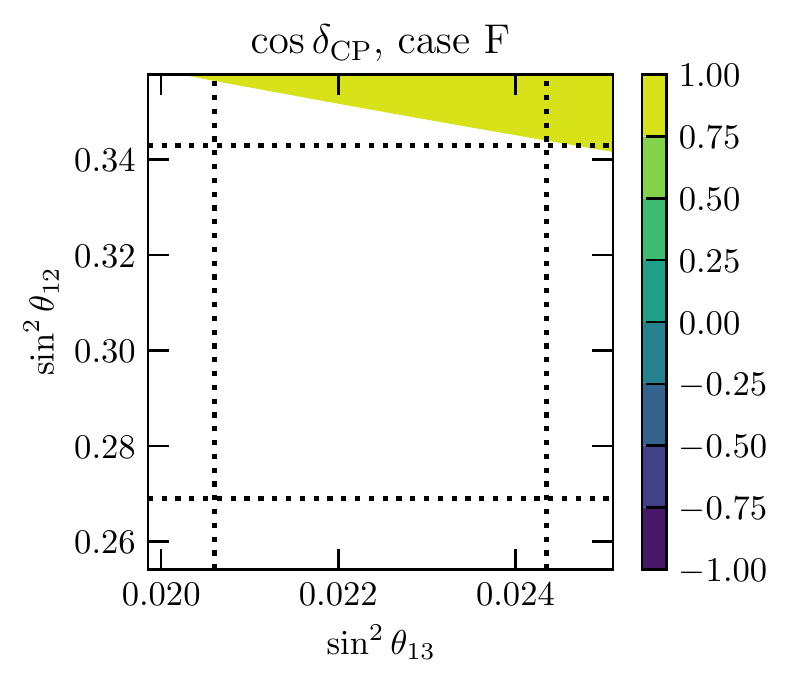}
    \caption{\label{fig:cos_d_EF}%
        Border of the physical region for $\cos\delta_\text{CP}$
        closest to the $\pm 3\sigma$ rectangle for $s_{13}^2$-$s_{12}^2$ (dotted lines).
        Predictions for $\cos\delta_\text{CP}$ use cases E (left) and F (right) from Ref.~\cite{Kang:2014mka}.
        White regions indicate unphysical $\cos\delta_\text{CP}$.
        }
\end{figure}

To provide an update for cases E and F,
we find that they cannot predict physical values for $\cos\delta_\text{CP}$
within the 3$\sigma$ ranges using current results from Ref.~\cite{Gonzalez-Garcia:2021dve}.
In Fig.~\ref{fig:cos_d_EF} the 3$\sigma$ rectangle
for $s_{13}^2$ and $s_{12}^2$ is shown
together with the closer physical boundary (colored contours)
for the predicted $\cos\delta_\text{CP}$ for both cases E and F.
Interestingly, in both panes of Fig.~\ref{fig:cos_d_EF},
the boundary of the physical predictions for $\cos\delta_\text{CP}$
is barely outside the 3$\sigma$ rectangle,
almost touching the upper right corner,
indicating that this level of exclusion must be quite recent.

With these results, all the cases with $U_0^\text{BM}$
considered in Ref.~\cite{Kang:2014mka}
can be considered ruled out at 3$\sigma$ or above.
Considering this, we will not follow the detailed analysis of the previous section
on the CP-violating Dirac phase for the cases of this section
and the rest of this work will be focused on cases A, B, C and D.

\section{Prospects at future experiments}
\label{sec:prospects}

To analyze the prospects for the four cases considered in this work,
we will employ simulations using the GLoBES software package~\cite{Huber:2004ka,Huber:2007ji}.
We will consider three long-baseline experiments: DUNE, T2HK and ESSnuSB\@.
For the DUNE experiment we consider the configuration
detailed in their technical design report~\cite{DUNE:2020ypp}.
According to Ref.~\cite{DUNE:2020ypp},
the DUNE experiment is planned to have a long-baseline of 1300~km,
with a 1.2~MW neutrino beam produced at Fermi National Accelerator Laboratory
and received at a far detector in Sanford Underground Research Facility.
This corresponds to $1.1\times 10^{21}$ protons on target (POT).
The far detector will consist of liquid argon time-projections chambers
and will have a (fiducial) mass of (40~kt) 70~kt.
In our simulation we assume a total run time of 7~years
equally distributed between neutrino and antineutrino modes.
In the case of the T2HK experiment we follow the setup described in Ref.~\cite{Hyper-Kamiokande:2016srs}.
A 1.3~MW neutrino beam will be produced at Japan Proton Accelerator Research Complex.
The neutrinos would arrive to a water Cherenkov detector
with a fiducial mass of 187~kt,
at a distance of 295~km.
A second identical detector is under consideration to be built in Korea.
Assuming a total of 10~years of operation of the first detector
it is possible to achieve $27\times 10^{21}$~POT\@.
Following Ref.~\cite{Hyper-Kamiokande:2016srs},
we assume that the 10~years run time is distributed
with a ratio of 3:1 for antineutrino to neutrino modes.
For ESSnuSB we consider the experimental setup outlined in Ref.~\cite{essnusb:2018}.
The neutrino beam would be produced at the European Spallation Source
with a power of 5~MW\@.
Neutrinos would be received at a MEMPHYS-like~\cite{MEMPHYS:2012bzz} water Cherenkov detector
with a (fiducial) mass of (507~kt) 1 Mt,
at a distance of 540~km.
With this configuration, ESSnuSB will reach $2.7\times 10^{23}$ POT per year.
In our analysis, we assume a run of 10~years
with a ratio of 8:2 for antineutrino to neutrino modes
as mentioned in the ``Nominal value'' column of Table~1.1 of Ref.~\cite{essnusb:2018}.

The statistical analysis follows the methodology described in Sec. III of Ref.~\cite{Blennow:2020ncm}.
To summarize the steps in this methodology,
we start with GLoBES $\chi_\text{G}^2$ function
comparing the $N^\text{obs}$ events observed in the simulation of the experiment
against $N^\text{th}$ events expected from theory.
GLoBES $\chi_G^2$ function can be written as
\begin{equation}
    \label{eq:chi2globes}
    \chi_\text{G}^2(\theta,\phi) = \sum_i\left[
        N_i^\text{th}(\theta,\phi) - N_i^{\rm obs}
        + N_i^{\rm obs} \ln\!\left(\frac{N_i^{\rm obs}}{N_i^\text{th}(\theta,\phi)}\right)
    \right]
\end{equation}
where $(\theta,\phi)$ refers to a set of parameters in the theory
and the summation run over bins.
Additionally, we include two Gaussian prior contributions to the total $\chi^2$
using the reported central values,
$s_{12,\text{obs}}^2$ and $s_{13,\text{obs}}^2$,
and their corresponding errors, $\sigma_{12}$ and $\sigma_{23}$,
given in Table~\ref{tab:nufit5_1_SK}~\cite{Gonzalez-Garcia:2021dve}.
Considering that currently the octant of $s_{23}^2$ is not known,
for its prior we use an interpolation of the $\chi^2$ table
provided in NuFIT's website~\cite{nufitwebsite}.
The full $\chi_\text{pr}^2$ is given by
\begin{equation}
    \label{eq:chi2prior}
    \chi_\text{pr}^2(\theta,\phi) = \left(\frac{s_{12}^2(\theta,\phi) - s_{12,\text{obs}}^2}{\sigma_{12}}\right)^2
        + \left(\frac{s_{13}^2(\theta,\phi) - s_{13,\text{obs}}^2}{\sigma_{13}}\right)^2
        + \chi^2_{23,\text{NuFIT}}(s_{23}^2(\theta,\phi)).
\end{equation}
The total $\chi^2$ to be minimized is given by
\begin{equation}
    \label{eq:chi2tot}
    \chi^2(\theta,\phi) = \chi_\text{G}^2(\theta,\phi) + \chi_\text{pr}^2(\theta,\phi).
\end{equation}
Following Ref.~\cite{Blennow:2020ncm}, our results will be presented for
$\Delta\chi^2 = \chi^2_\text{mod} - \chi^2_\text{free}$,
where $\chi^2_\text{mod}$ is the result of minimizing Eq.~\eqref{eq:chi2tot}
over the model parameters $\theta$ and $\phi$,
while $\chi^2_\text{free}$ is the minimization over oscillation parameters
ignoring constraints from the scenarios of Sec.~\ref{sec:pertTBM}.

\begin{figure}[htpb]
    \includegraphics{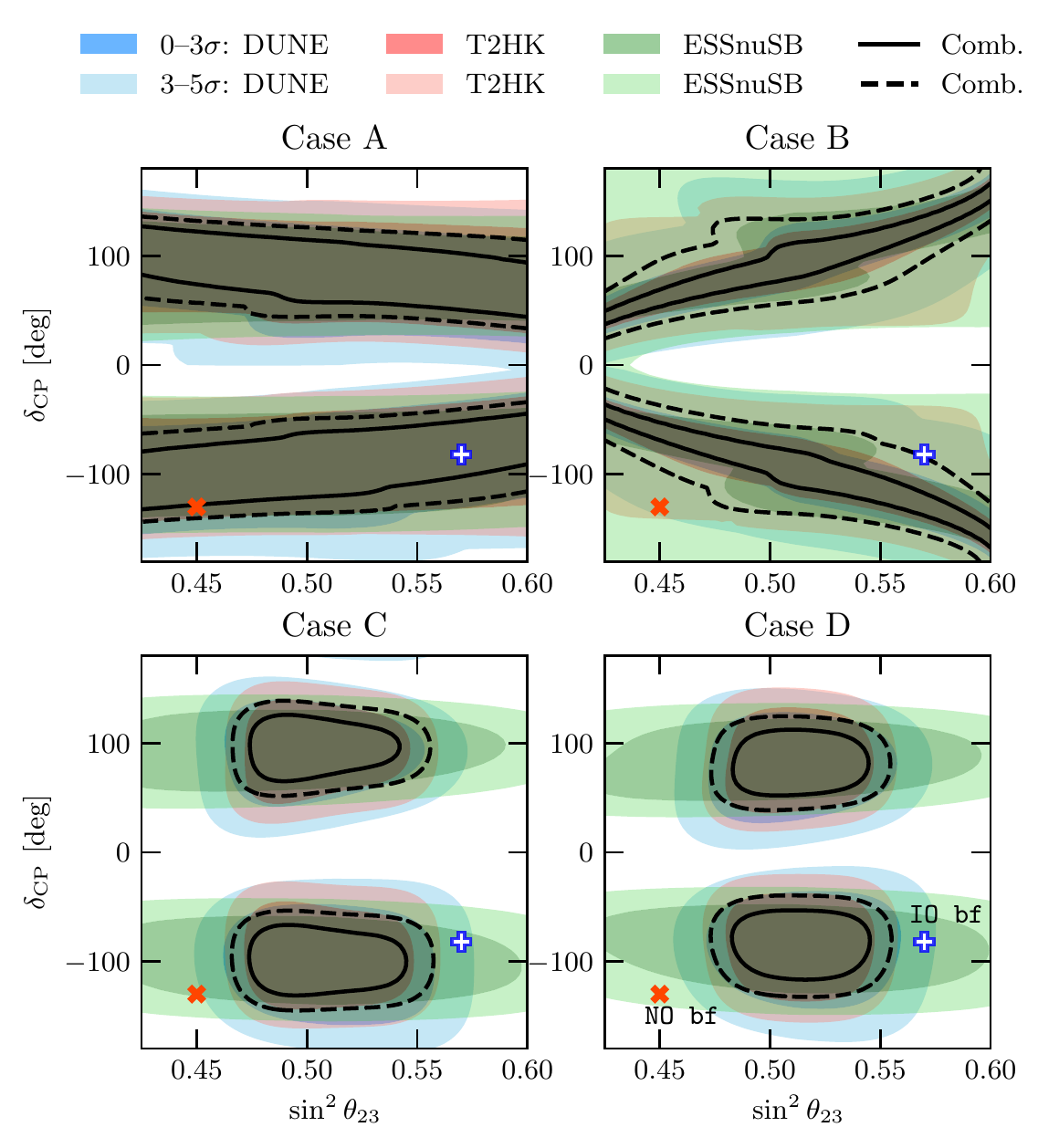}
    \caption{\label{fig:simulations}%
        Prospects of future experiments excluding cases A, B, C and D
        in the plane $\sin^2\theta_{23}$-$\delta_\text{CP}$.
        A measurement in the white region (outside dashed black contour)
        indicates that the corresponding case could be excluded
        with $5\sigma$ or more confidence by the experiment (combined experiments).
        A measurement in the light colored region (between solid and dashed black contour)
        indicates an exclusion between 3 to 5$\sigma$.
        If the experiment (combined experiments) measures a true value
        inside the darker region (solid black contour)
        then the result and the model are compatible within 3$\sigma$.
        The experimental results used in the simulation correspond to normal ordering,
        however, there is no significant change for inverted ordering
        other than the current best fit point, indicated as a red thick $\times$ for NO
        and a blue-white + for IO.}
\end{figure}

The results of our simulations are presented in Fig.~\ref{fig:simulations}\@.
We performed scans over the true values in the plane $s_{23}^2$-$\delta_\text{CP}$,
while fixing other true values to their central values, given in Table~\ref{tab:nufit5_1_SK}.
One obvious feature is that cases C and D have a more constrained $s_{23}^2$
compared to cases A and B\@.
This is expected from the fact that, in cases C and D,
$s_{23}^2$ depends on the value of $s_{13}^2$
which reduces its allowed range,
while for cases A and B $s_{23}^2$ is free.
For cases C and D, the compatible $s_{23}^2$ is more strongly constrained by DUNE and T2HK,
while for all cases ESSnuSB reduces the $\delta_\text{CP}$ phase range.
Assuming that future experimental results
will be close to the current best fit point (NO\@: red thick $\times$, IO\@: blue-white +),
we can see that T2HK (red regions) alone
could exclude cases C and D for both NO and IO at 5$\sigma$ or more,
while DUNE (blue regions) could exclude C and D only for the NO result,
with the IO result remaining within 3 to 5$\sigma$.
Under the same assumed future results, ESSnuSB could not exclude any case above 5$\sigma$.
However, the combination of the three experiments (black contours)
has the capacity of excluding cases B, C and D
for both orderings to 5$\sigma$ or more.
Case A has the best chances of survival, with a NO result disfavoured only between 3 to $\sigma$
and IO staying well below 3$\sigma$.

\section{A model with \texorpdfstring{$A_4$}{A\textfourinferior} modular symmetry}
\label{sec:modularA4}

In this section we will construct a model
that predicts the neutrino masses and mixing
within the measured limits,
and we will show that symmetry breaking in this model
results in a mixing pattern that is consistent
with case A studied in previous sections.

The properties of modular forms are described in detail in Ref.~\cite{Feruglio:2017spp}.
To summarize the modular approach to flavor models,
consider the group $\Gamma(N)$ defined by
\begin{equation}
	\Gamma(N) = \left\{
		\begin{pmatrix}
			a & b \\
			c & d
		\end{pmatrix}
		\in SL(2, Z),
		\begin{pmatrix}
			a & b \\
			c & d
		\end{pmatrix} =
		\begin{pmatrix}
			1 & 0 \\
			0 & 1
		\end{pmatrix}
		({\rm mod}\ N)
	\right\},
\end{equation}
where $SL(2,Z)$ is the special linear group of 2$\times$2 matrices
with integer elements and determinant equal to 1.
The elements of the group $\Gamma(N)$ transform a complex variable $\tau$, 
constrained by ${\rm Im}(\tau) > 0$, according to
\begin{equation}
    \gamma\tau = \frac{a\tau + b}{c\tau + d}
\end{equation}
we call this a linear fractional transformation.
The group of these linear fractional transformations,
called the modular group $\overline{\Gamma}(N)$,
is related to $\Gamma(N)$:
for $N \leq 2$, $\overline\Gamma(N)\equiv \Gamma(N)/\{\pm\mathds{1}\}$,
while for $N>2$ we have $\overline\Gamma(N)\equiv \Gamma(N)$.
The generators of the group $\overline\Gamma$ can be expressed
using the $SL(2,Z)$ matrices
\begin{equation}
    \label{eq:overGammaST}
		S = \begin{pmatrix}
			0 & -1 \\
			1 & 0
		\end{pmatrix},\quad
		T = \begin{pmatrix}
			1 & 1 \\
			0 & 1
		\end{pmatrix}
\end{equation}
which satisfy the relation $S^2 = (ST)^3 = \mathds{1}$.
The quotient $\overline\Gamma/\overline\Gamma(N)$
defines finite groups referred as finite modulars groups $\Gamma_N$.
The generators of these groups have the additional property that $T^N=\mathds{1}$.
For $N\in \{2, 3, 4, 5\}$ these groups are isomorphic
to the permutation groups $S_3$, $A_4$, $S_4$ and $A_5$, respectively.
For further details on modular forms
and their relation to the permutation groups mentioned,
the interested reader may check Refs.~\cite{Feruglio:2017spp,Novichkov:2019sqv,Kobayashi:2022moq}.

The model that we develop here is based on modular forms of level $N=3$
which have a quotient group, $\Gamma_3$, isomorphic to $A_4$,
the symmetry group of the tetrahedron.
For $A_4$ the generators have the properties
\begin{equation}
    S^2 = (ST)^3 = T^3 = \mathds{1}.
\end{equation}
The modular forms of level 3 were constructed on Appendix C of Ref.~\cite{Feruglio:2017spp}
and correspond to
\begin{align}
	Y_1(\tau) & = \frac{i}{2\pi}\left[
		\frac{\eta'\left( \frac{\tau}{3} \right)}{\eta\left( \frac{\tau}{3} \right)}
		+\frac{\eta'\left(\frac{\tau + 1}{3}\right)}{\eta\left(\frac{\tau + 1}{3}\right)}
		+\frac{\eta'\left(\frac{\tau + 2}{3}\right)}{\eta\left(\frac{\tau + 2}{3}\right)}
		- \frac{27\eta'(3\tau)}{\eta(3\tau)}
	\right],\nonumber\\
    \label{eq:A4modforms}
	Y_2(\tau) & = \frac{-i}{\pi}\left[
		\frac{\eta'\left( \frac{\tau}{3} \right)}{\eta\left( \frac{\tau}{3} \right)}
		+\omega^2 \frac{\eta'\left(\frac{\tau + 1}{3}\right)}{\eta\left(\frac{\tau + 1}{3}\right)}
		+\omega\frac{\eta'\left(\frac{\tau + 2}{3}\right)}{\eta\left(\frac{\tau + 2}{3}\right)}
	\right],\\
	Y_3(\tau) & = \frac{-i}{\pi}\left[
		\frac{\eta'\left( \frac{\tau}{3} \right)}{\eta\left( \frac{\tau}{3} \right)}
		+\omega \frac{\eta'\left(\frac{\tau + 1}{3}\right)}{\eta\left(\frac{\tau + 1}{3}\right)}
		+\omega^2\frac{\eta'\left(\frac{\tau + 2}{3}\right)}{\eta\left(\frac{\tau + 2}{3}\right)}
	\right],\nonumber
\end{align}
where $\eta$ is the Dedekind eta function defined by
\begin{equation}
    \eta(\tau) = q^{1/24} \prod_{n=1}^\infty (1 - q^n)\,, \quad q\equiv \exp(i2\pi \tau)\,,\quad {\rm Im}(\tau) > 0\,,
\end{equation}
and $\omega = (-1 + i\sqrt{3})/2$.
The forms $Y_i$ belong to a $A_4$ triplet $(Y_1, Y_2, Y_3)\equiv Y$.


\subsection{Lepton masses}

\begin{table}[tb]
    \setlength\tabcolsep{0.1cm}
    \begin{tabular}{lcccccccccc}
        \toprule
                & $e^c$, & $\mu^c$, & $\tau^c$ & $N^c$ & $L_e$, & $L_\mu$, & $L_\tau$ & $H_d$ & $H_u$ & $\chi$   \\
                \cmidrule(lr){2-4} \cmidrule(lr){5-5} \cmidrule(lr){6-8} \cmidrule(lr){9-9} \cmidrule(lr){10-10} \cmidrule(lr){11-11}
        $SU(2)_L\times U(1)_Y$ & \multicolumn{3}{c}{$(1, +1)$} & $(1,0)$  & \multicolumn{3}{c}{$(2,-1/2)$} & $(2,-1/2)$ & $(2,+1/2)$ & $(1,0)$ \\
        $A_4$    & $\mathbf{1}$, & $\mathbf{1}''$, & $\mathbf{1}'$ & $\mathbf{3}$ & $\mathbf{1}$, & $\mathbf{1}'$, & $\mathbf{1}''$ & $\mathbf{1}$ & $\mathbf{1}$ & $\mathbf{1}$ \\
        $U(1)_X$ & $-\frac{1}{2}-f_e$, & $-\frac{1}{2}-f_\mu$, & $-\frac{1}{2}-f_\tau$ & $-\frac{1}{2}$ & \multicolumn{3}{c}{$\frac{1}{2}$} & 0 & 0 & 1 \\
        $k_I$   & \multicolumn{3}{c}{$4$} & $2$ & \multicolumn{3}{c}{0} & 0 & 0 & 0 \\
        \bottomrule
    \end{tabular}
    \caption{\label{tab:reps_f}%
        Fields of the model, their representation under the symmetries considered
        and modular weights $k_I$.
        }
\end{table}

We will consider Majorana neutrinos
that acquire small masses via the see-saw mechanism.
This model is based on an extension by the symmetry group $A_4\times U(1)_X$
with the right handed neutrinos in a triplet of chiral supermultiplets $N^c$.
The field content of the model, $A_4$ representation, $U(1)_X$ charges and modular weights $k_I$
are collected in Table~\ref{tab:reps_f}\@.
With those assignments for the fields, the superpotential is given by
\begin{align}
    \label{eq:superpot}
    W = {}& \alpha_1\,e^c L_e Y^{(4)}_\mathbf{1} \Big(\frac{\chi}{\Lambda}\Big)^{f_e} H_d
            + \alpha_2\,\mu^c L_\mu Y^{(4)}_\mathbf{1} \Big(\frac{\chi}{\Lambda}\Big)^{f_\mu} H_d
            + \alpha_3\,\tau^c L_\tau Y^{(4)}_\mathbf{1} \Big(\frac{\chi}{\Lambda}\Big)^{f_\tau} H_d\nonumber\\
          & + \beta_1(N^c Y)_\mathbf{1} L_e H_u
            + \beta_2(N^c Y)_{\mathbf{1}''} L_\mu H_u
            + \beta_3(N^c Y)_{\mathbf{1}'} L_\tau H_u\nonumber\\
          & + \gamma_1(N^c N^c)_\mathbf{1}Y^{(4)}_\mathbf{1}\chi
            + \gamma_2(N^c N^c)_{\bf 3}Y^{(4)}_{\bf 3}\chi\,,
\end{align}
where $\alpha_i$, $\beta_i$ and $\gamma_i$ are dimensionless couplings.
In the case of $\alpha_i$ and $\beta_i$ they can be made real by field redefinitions
and are taken as $\mathcal{O}(1)$ coefficients.
The $\gamma_i$ couplings are complex
and we take their modulus as $\mathcal{O}(1)$.
The weight 4 modular forms are given by
\begin{align}
    Y_\mathbf{1}^{(4)} & = Y_1^2 + 2 Y_2 Y_3, \\
    Y_\mathbf{3}^{(4)} & = (Y_1^2 - Y_2 Y_3, Y_3^2 - Y_1 Y_2, Y_2^2 - Y_1 Y_3).
\end{align}

At energies below the electroweak scale,
the scalars $\chi$, $H_u$ and $H_d$ acquire vacuum expectation value (VEV)
giving masses to the fields in the superpotential of Eq.~\eqref{eq:superpot}.
The charged lepton masses can be extracted from the first line
of the superpotential and correspond to the diagonal matrix
\begin{equation}
    {\cal M}_\ell = Y^2_1 \langle H_d\rangle(1+2ab)\,{\rm diag}\left(
        \alpha_1\left(\frac{\langle\chi\rangle}{\Lambda}\right)^{f_e},
        \alpha_2\left(\frac{\langle\chi\rangle}{\Lambda}\right)^{f_\mu},
        \alpha_3\left(\frac{\langle\chi\rangle}{\Lambda}\right)^{f_\tau}
    \right),
\end{equation}
where $a \equiv Y_2/Y_1$ and $b\equiv Y_3/Y_1$.
One can choose integers $f_{e,\mu,\tau}$ and $\langle\chi\rangle/\Lambda$
in order for $(\langle\chi\rangle/\Lambda)^{f_e-f_\tau}=0.0003$ and $(\langle\chi\rangle/\Lambda)^{f_\mu-f_\tau}=0.06$
to satisfy the empirical results of charged lepton masses.
From the second and third lines in the superpotential
we can read off the following mass matrices
\begin{align}
    m_D & = Y_1\langle H_u\rangle
    \begin{pmatrix}
        \beta_1  & \beta_2 b & \beta_3 a  \\
        \beta_1 b& \beta_2 a & \beta_3  \\
        \beta_1 a & \beta_2 & \beta_3 b
    \end{pmatrix},\\
    M_R & = Y^2_1\langle\chi\rangle\gamma_1
    \begin{pmatrix}
        1 + \frac{4}{3}\gamma + a b \left(2-\frac{4}{3}\gamma\right) & 2 \gamma b & -\frac{2}{3}\gamma\left(b^2 - a\right)  \\
        (M_R)_{1,2} & \frac{4}{3}\gamma\left(b^2 - a\right) & 1 - \frac{2}{3}\gamma + a b \left(2+\frac{2}{3}\gamma\right)  \\
        (M_R)_{1,3} & (M_R)_{2,3} & -4\gamma b
    \end{pmatrix},
    \label{eq:nuMassMatrix}
\end{align}
respectively,
with $\gamma \equiv \gamma_2/\gamma_1$.
The matrix $m_D$ corresponds to the Dirac masses for the neutrinos
and the symmetric matrix $M_R$ is for the Majorana masses of the right handed neutrinos.
From these matrices we obtain the light neutrino mass matrix
\begin{align}
    \label{eq:lightNuMassMatrix}
    {\cal M}_\nu  &= - m^T_D M^{-1}_R m_D \,.
\end{align}
Note that the light mass matrix is proportional to $\langle H_u\rangle^2/\langle \chi\rangle$,
therefore, neutrino masses are expected to be small for very large $\langle \chi \rangle$.

Since the $A_4$ flavor models of leptons
give large flavor mixing angles clearly~\cite{Ma:2001dn,Babu:2002dz},
several $A_4$ modular invariant models have been proposed~\cite{Criado:2018thu,Kobayashi:2018scp,Ding:2019zxk,Zhang:2019ngf,Okada:2020brs}.
It may be useful to comment on the distinctive features of our model.
Our charged lepton mass matrix is diagonal,
in contrast to previous models,
by assignment of $A_4$ singlets for both left-handed and right-handed leptons apart from the right-handed neutrinos.
Then, the lepton mixing angles come from
flavor structure of the neutrino mass matrices.
Therefore, our model is advantageous for discussing the TBM and the case A
in the context of $A_4$ flavor symmetry.
It is emphasized that the Dirac phase $\delta_\text{CP}\approx 3/2\pi$
could be reproduced
around the fixed point $\tau=i$ as seen in the next subsection.

\subsection{Perturbative modifications to TBM mixing}

\begin{table}[tb]
    \setlength\tabcolsep{0.2cm}
    \begin{tabular}{lccccc}
        \toprule
             &        $\tau$         & $\beta_1$ & $\beta_2$ & $\beta_3$ &       $\gamma$       \\
        \midrule
        A1   & $ 0.30607 + i 0.96354$ &   1.3036  &  1.4064   &  1.6484   & $1.1647 + i 0.30861$ \\
        TBM1 & $-0.16155 + i 0.99335$ &  0.072040 &  1.4753   &  1.6665   & $1.6552 + i 0.18674$ \\
        \midrule
        A2   & $ 0.30311 + i 0.97203$ &   1.1102  &  1.2924   &  1.5131   & $1.0273 + i 0.32692$ \\
        TBM2 & $-0.15361 + i 1.00429$ &  0.080130 &  1.3573   &  1.5345   & $1.6377 + i 0.46016$ \\
        \midrule
        A3   & $ 0.31013 + i 0.95205$ &   1.2011  &  1.2952   &  1.5853   & $1.0030 + i 0.40452$ \\
        TBM3 & $-0.16801 + i 0.99657$ &  0.078129 &  1.3847   &  1.5797   & $1.4620 + i 0.31883$ \\
        \bottomrule
    \end{tabular}
    \caption{\label{tab:A4Params}%
        Benchmark points for the model presented in Sec.~\ref{sec:modularA4}
        that predict a pattern consistent with case A
        and a neighboring point that predicts TBM mixing.
        }
\end{table}
\begin{table}[tb]
    \setlength\tabcolsep{0.2cm}
    \begin{tabular}{lcccccc}
        \toprule
           & $s_{12}^2$ & $s_{13}^2$ & $s_{23}^2$ & $\delta_\text{CP}$ [deg] & $\Delta m_{21}^2$ [eV$^2$] & $\Delta m_{3k}^2$  [eV$^2$] \\
        \midrule
        A1 &    0.318   &   0.02243  &    0.448   &            257           &    $7.47\times 10^{-5}$    &    $2.514\times 10^{-3}$    \\
        A2 &    0.3182  &   0.02224  &    0.450   &            257           &    $7.41\times 10^{-5}$    &    $2.515\times 10^{-3}$    \\
        A3 &    0.318   &   0.02242  &    0.426   &            251           &    $7.52\times 10^{-5}$    &    $2.495\times 10^{-3}$    \\
        \bottomrule
    \end{tabular}
    \caption{\label{tab:A4Predictions}%
        Predictions for the oscillation parameters
        using the corresponding point from Table.~\ref{tab:A4Params}.}
\end{table}

As mentioned in Sec.~\ref{sec:pertTBM},
in the cases where the perturbation to TBM mixing
is due to the breaking of residual symmetries in the neutrinos sector,
such as the model presented here,
we can expect perturbations of the form of cases A or B.

Considering the constraints imposed on $s_{12}^2$
and mentioned before the start of Sec.~\ref{sec:probDensCosDelta},
case B is unable to reproduce the current best fit value for $s_{12}^2$.
This leaves case A as the most appropriate candidate
for realistic phenomenological studies.
Here we will attempt to show that the $A_4$ model presented above
can predict oscillation parameters that are consistent with case A
and are in complete agreement with the current best fit limits
summarized in Table~\ref{tab:nufit5_1_SK}.

As a first step, we find a few parameter choices
that give predictions with good agreement with current experimental values
and are consistent with case A, characterized by Eqs.~(\ref{eq:A12delCP}).
We provide a few benchmark points in Table~\ref{tab:A4Params}
labeled as A$j$ as well as their predictions in Table~\ref{tab:A4Predictions}.
The next step is finding a neighboring point
that reproduces TBM with good accuracy.
Such point should be considered only illustrative,
since TBM is in disagreement with current bounds,
namely with the measured range for $s_{13}^2$.
The TBM points neighboring the A$j$ points
are given in Table~\ref{tab:A4Params} with the label TBM$j$.
Finally, we can compare these two types of points
to assess how much each parameter changes
between TBM and the perturbed case A.

We found that TBM mixing, particularly $s_{13}^2 = 0$,
would require $\beta_1 \approx 0$.
Getting the other $s_{ij}^2$ requires $\tau \approx -0.16 + i 1.0$
and $\beta_2/\beta_3 \approx 0.9$.
Note that the points TBM$j$ in Table~\ref{tab:A4Params} do not use the value
$\beta_1 = 0$ since it would make the first column of $m_D$ exactly zero.
The overall factor of $\mathcal{M}_\nu$ is taken as
$\langle H_u \rangle^2/\langle \chi \rangle|\gamma_1| = 1.0092\times 10^{-11}$~GeV\@.
When comparing the points in Table~\ref{tab:A4Params}
we see that $\beta_1$ and $\beta_3$ change the least
remaining identical to 2 significant digits.
Expectedly, $\beta_1$ changes the most
since this parameter is related to the appearance of non-zero $\theta_{13}$.
The parameter $\tau$ is dominated by its imaginary part
which changes roughly 3\%.
The parameter $\gamma$ is dominated by its real part
and is the one that changes the most,
and is mostly related to the requirement that the A$j$ points
predict appropriate mass differences
which we do not require from the TBM points for simplicity.

To conclude this section with a comment,
while the $A_4$ model presented above permits mixing patterns far more complicated,
the study of this section illustrates
how case A may arise in a realistic model.
Moreover, the relation that exists between case A and TBM mixing
is made explicit in the comparison between model parameter values.
This analysis is independent of the model
and could be an starting point for a detailed study of the effects of breaking the residual symmetries
that led to TBM mixing in the first place.

\section{Conclusion}
\label{sec:conclusion}

In this work we revisit the perturbed mixing patterns
that were considered in Ref.~\cite{Kang:2014mka}
for the popular BM and TBM mixings.
Using current best fit values and 3$\sigma$ ranges
for the oscillation parameters
we found that the considered perturbations to BM mixing,
labeled E and F, cannot predict physical values
for $\cos\delta_\text{CP}$ with $s_{12}^2$ and $s_{13}^2$ inside their 3$\sigma$ ranges,
while the four cases that consider perturbations
of TBM mixing survived.
We extended on previous efforts to predict the leptonic CP-violating Dirac phase
by calculating distributions for its allowed values
in light of the relations between oscillations parameters.
For cases A, B and C we found that the preferred $\delta_\text{CP}$ phase
is located around 270$^\circ$
while for case B the most favoured values spanned a range
roughly from 200$^\circ$ to 320$^\circ$.
These values consider that, according to Ref.~\cite{Gonzalez-Garcia:2021dve},
the observed preferred range for $\delta_\text{CP}$ is between 144$^\circ$ and 350$^\circ$.
Interestingly, planned experiments will have the power to constrain these simple perturbations,
particularly cases B, C and D, which have the most constraining conditions.
In the case of B, $s_{12}^2 > 1/3$ is in tension with the currently measured value,
and if future experiments keep this tendency we will see the tension increased.
For cases C and D, due to each case predicting $s_{23}^2$ in different octants,
one of them will be excluded when the octant problem is resolved.
Nonetheless, both cases, C and D, predict $s_{23}^2$ quite close to 1/2,
and if $s_{23}^2$ stays in close proximity to its current central value
both cases could eventually be ruled out.
The simulations performed and described in Sec.~\ref{sec:prospects}
show that DUNE2, T2HK and ESSnuSB experiments have the combined capacity
to rule out cases B, C and D by more than 5$\sigma$,
while case A could be left disfavoured by more than 3$\sigma$.
We finalize by showing the emergence of case A
from an $A_4$ modular symmetry flavor model.
This model is capable of predicting currently measured oscillation parameters
within their acceptable ranges.
Moreover, we showed the existence of nearby points that predict TBM mixing
to illustrate the degree of perturbation in the parameters
required to obtain the mixing pattern of case A\@.
The results of this study can be applied to any model
that results in a mixing pattern consistent with the list in Eq.~\eqref{eq:cases}.
Furthermore, any of the steps performed in this study could be applied
to different neutrino masses and mixing models
for which one can obtain relations like those in Eqs.~\eqref{eq:A12delCP} to~\eqref{eq:D12delCP},
and may help reveal details brought about by the existence of such constraints.

\end{document}